\documentclass{article}
\usepackage[top=1in,bottom=1in,left=1in,right=1in]{geometry}  

\usepackage[utf8]{inputenc}
\usepackage{graphicx, xcolor, amsmath}
\usepackage{setspace}
\usepackage[colorlinks,allcolors=cyan!70!black]{hyperref}

\def\fref#1{{Figure~\ref{#1}}}

\newcommand{\suppfref}[1]{{Figure~\ref{#1}}}

\usepackage{enumitem}

\newcommand{\mathbbm}[1]{\text{\usefont{U}{bbm}{m}{n}#1}}
\newcommand\corrauthor[1]{%
  \begingroup
  \renewcommand\thefootnote{}\footnote{#1}%
  \addtocounter{footnote}{-1}%
  \endgroup
}

\begin{document}

\begin{center}
\begin{spacing}{1.5}
\textbf{\Large EGGTART: A computational tool to visualize the dynamics of biophysical transport processes
under the inhomogeneous $\ell$-TASEP}
\end{spacing}

\vspace{5mm}
Dan D. Erdmann-Pham$^{1,*}$, Wonjun Son$^{2.*}$\corrauthor{$^*$ equal contribution}, Khanh Dao Duc$^{3,**}$ and Yun S. Song$^{4,5,6,**}$ \corrauthor{$^{**}$ corresponding authors: kdd@math.ubc.ca, yss@berkeley.edu}

\vspace{5mm}
$^{1}$Department of Mathematics, University of California, Berkeley, CA 94720, USA\\
$^{2}$Department of Computer Science, Columbia University, New York, NY 10027, USA\\
$^{3}$Department of Mathematics, University of British Columbia, Vancouver, BC V6T 1Z4, Canada\\
$^{4}$Department of Statistics, University of California, Berkeley, CA 94720, USA\\
$^{5}$Computer Science Division, University of California, Berkeley, CA 94720, USA\\
$^{6}$Chan Zuckerberg Biohub, San Francisco, CA 94158, USA

\vspace{5mm}
\today
\vspace{5mm}
\end{center}

\begin{abstract}

The totally asymmetric simple exclusion process (TASEP), which describes the stochastic dynamics of interacting particles on a lattice, has been actively studied over the past several decades and applied to model  important biological transport processes.  Here we present a software package, called EGGTART (\textbf{E}xtensive \textbf{G}UI \textbf{g}ives \textbf{TA}SEP-realization in \textbf{r}eal \textbf{t}ime), which quantifies and visualizes the dynamics associated with a generalized version of the TASEP with an extended particle size and heterogeneous jump rates. This computational tool is based on analytic formulas obtained from deriving and solving the hydrodynamic limit of the process. It allows an immediate quantification of the particle density, flux, and phase diagram, as a function of a few key parameters associated with the system, which would be difficult to achieve via conventional stochastic simulations.  Our software should therefore be of interest to biophysicists studying general transport processes, and can in particular be used in the context of gene expression to model and quantify mRNA translation of different coding sequences.
\end{abstract}

\section*{Introduction}

The totally asymmetric exclusion process (TASEP) is a stochastic process used to model a large variety of transport phenomena involving interacting particles \cite{schadschneider2010}. Although it has been studied extensively over the past several decades, it is still the subject of active research and is replete with many open problems motivated by biophysical applications. In particular, the recent emergence of experimental data through advances in microscopy and sequencing techniques has revealed a wealth of molecular processes that are well-described by TASEP models, including the motion of ribosomes \cite{korkmazhan2017dynamics, szavits2018deciphering}, 
RNA polymerase
\cite{van2017crowding},
and motor proteins
\cite{miedema2017correlation}.
Such processes require extending the classical TASEP by several layers of complexity, 
and as a result, their analysis 
has remained elusive and restricted to 
particular cases 
that do not adequately reflect biology. This shortcoming recently led us to study
the so-called inhomogeneous $\ell$-TASEP, where particles of size $\ell$ traverse a lattice of inhomogeneous jump rates. By considering the  hydrodynamic limit of the process, we
obtained exact formulas for particle currents and densities~\cite{erdmann2020key}, offering 
immediate quantification of the particle density, flux, and phase diagram, which would be difficult to achieve via conventional stochastic simulations.

To help to visualize these theoretical results, we present here a software package called EGGTART (\textbf{E}xtensive \textbf{G}UI \textbf{g}ives \textbf{TA}SEP-realization in \textbf{r}eal \textbf{t}ime). 
For an arbitrary specification of the key system parameters (particle size, entrance rate, exit rate, and site-specific jump rates), EGGTART provides a graphical user interface that allows to extract the main quantities of interest (e.g., the current and local densities of particles), in addition to providing a phase diagram that fully describes different types of traffic behavior.  We demonstrate how it can recover various versions of the TASEP studied in the past literature as special cases. By facilitating the exploration and visualization of both theoretical and practical aspects of the inhomogeneous $\ell$-TASEP, 
we anticipate our computational tool to be broadly helpful to mathematicians, physicists and biologists in studying various biophysical transport phenomena.

\section*{Methods}
\subsection*{The inhomogeneous $\ell$-TASEP}

\begin{figure}[h]
    \centering
    \includegraphics[trim = 0mm 0mm 0mm 3mm, clip,width=0.35\textwidth]{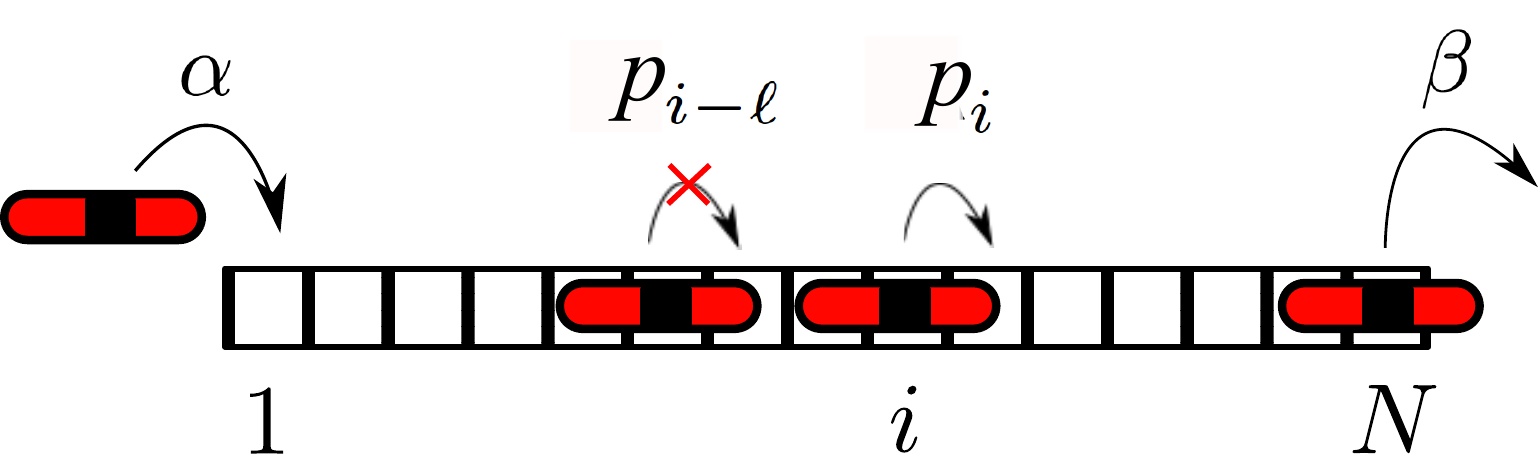}
    \caption{\textbf{Illustration of the inhomogeneous $\ell$-TASEP with open boundaries.} Particles (of size $\ell=$ 3 here) enter
the first site of the lattice at rate $\alpha$ and a particle at position
$i$ (here denoted by the position of the midpoint of the particle)
moves one site to the right at rate $p_i$, provided that the next
$\ell$ sites are empty. A particle at the end of the lattice exits 
at rate $\beta$. }
    \label{fig_LTASEP} 
\end{figure}

The inhomogeneous $\ell$-TASEP is a Markov process, illustrated in \fref{fig_LTASEP}, where particles of size $\ell$ jump unidirectionally along a lattice of $N$ sites under mutual exclusion: a particle at site $i \in \{1, \dots, N-1\}$ remains at its position if the 
site $i+\ell$ 
is occupied, and jumps at exponential rate $p_i$ to position $i+1$ otherwise. Particles can enter the lattice at the first site at rate $\alpha$ 
while respecting the mutual exclusion constraint and leave the lattice at the last site 
at rate $\beta$.

\subsection*{The hydrodynamic limit}

Although solving the inhomogeneous $\ell$-TASEP is in general intractable, we can do so in the continuum limit of $N\to\infty$. More precisely, denoting by $\tau(t) \in \{0,1\}^N$ a configuration at time $t$ and assuming that $\lambda(x) = \lim_{N\to\infty}
\frac{1}{\ell}\sum_{i=0}^{\ell-1}p_{\lfloor Nx \rfloor+i}$ exists and is differentiable, we have shown in \cite{erdmann2020key} that the $\ell$-smoothed density of particles $\rho(x,t)\mathrm{d}x = \lim_{N\to\infty} \frac{1}{N}\sum_{n=1}^{N-\ell+1} \frac{1}{\ell} \sum_{i=0}^{\ell-1} \tau_{n+i}(Nt) \delta_{\frac{n+i}{N}} (\mathrm{d}x) 
$, satisfies the inhomogeneous conservation law
\begin{equation}
    \partial_t \rho = -\partial_x\left[\lambda(x) \rho\, G(\rho) \right],
    \label{eq:hydro_limit} 
\end{equation}
where $G(\rho) = \frac{1- \ell \rho}{1- (\ell-1)\rho}$.
For constant jump rates and $\ell = 1$
equation \eqref{eq:hydro_limit} reduces to the
one-dimensional Burgers' equation, known as the hydrodynamic limit of the classical TASEP \cite{blythe2007nonequilibrium}. As shown next, increasing $\ell$ and
introducing spatial inhomogeneity 
lead to more complex solutions.

\subsection*{Analytical solutions and phase diagram}\label{sec:phase_diagram}

Equation~\eqref{eq:hydro_limit} can be solved using the method of characteristics \cite{erdmann2020key}, wherein the two boundary points $0$ and $1$ each emit a characteristic curve whose evolution determines the long-term behavior of the system. 
These curves are controlled 
by boundary conditions, which 
yield a phase diagram in $\alpha$ and $\beta$. 
Surprisingly, this phase diagram is completely characterized by four parameters: the particle size $\ell$, and the minimal, initial and terminal jump rates $\lambda_{\min} := \min_x \lambda(x), \lambda_0 := \lambda(0)$ and $\lambda_1 := \lambda(1)$, respectively. These parameters determine the critical rates $\alpha^{\ast} = \alpha^{\ast}(\ell, \lambda_0, \lambda_{\min})$ and $\beta^{\ast} = \beta^{\ast}(\ell, \lambda_1, \lambda_{\min})$ at which phase transitions occur (precise expressions are provided in \cite{erdmann2020key}). The different regions of the phase diagram 
are characterized as: 

\begin{enumerate}[itemsep=5pt,parsep=0pt]
    \item $\alpha < \alpha^{\ast}$ and $\beta > \beta^{\ast}$: In this \textit{low density regime} (LD), the low entrance rate coupled with large exit rate establish an overall small density $\rho_L(x) = \rho_L(\lambda(x),\ell,J_L) < \rho^{\ast} := (\ell + \sqrt{\ell})^{-1}$, where $J_L = \frac{\alpha(\lambda_0 - \alpha)}{\lambda_0 + (\ell-1)\alpha}$ is the particle current. 
    \item  $\alpha > \alpha^{\ast}$ and $\beta < \beta^{\ast}$: Particles are injected frequently and drained slowly, shaping the \textit{high density phase} (HD). The associated density $\rho_R(x) = \rho_R(\lambda(x),\ell, J_R) > \rho^{\ast}$ is strictly larger than $\rho_L$, with a current now given by $J_R = \frac{\beta(\lambda_1 - \beta)}{\lambda_1 + (\ell-1)\beta}$.
    \item $\alpha < \alpha^{\ast}$ and $\beta < \beta^{\ast}$: A phase transition occurs along the non-linear curve $J_L = J_R$, with $J_L < J_R$ resulting in LD densities and currents, and $J_L > J_R$ in HD ones.
    \item $\alpha > \alpha^{\ast}$ and $\beta > \beta^{\ast}$: This phase characterizes the \textit{maximum current regime} (MC). The maximal current $J_{\max} = \lambda_{\min}\cdot\big(1+\sqrt{\ell} \big)^{-2}$ results from of superposition of high and low densities $\rho(x) = \rho_R(x)\cdot \mathbbm{1}_{x \leq x_{\min}} + \rho_L(x) \cdot \mathbbm{1}_{x \geq x_{\min}}$, where $x_{\min} = \arg\min_x \lambda(x)$. 
\end{enumerate}

\subsection*{General description of the software and availability}

 For any parametrization of the inhomogneous $\ell$-TASEP (e.g., particle size, entrance, exit and site-specific rates), our computational tool EGGTART integrates the analytical formulas given above into a graphical user interface that 
 provides a quantification and visualization of key properties, 
 such as the particle  current, densities, and phase diagram associated with different types of particle traffic behavior.
It has been developed in Python~3.6 and uses the \texttt{numpy} and \texttt{pyqtgraph}  packages.
Versions for Mac OS~X, Linux (Ubuntu), and Windows (10) are available at \url{https://github.com/songlab-cal/eggtart}, together with a user manual describing
general features, and a tutorial
to be used with the demo .csv input files.

\section*{Results}
We demonstrate the utility of EGGTART for a wide range of applications and examples, ranging in order of increasing complexity from the original TASEP model to the general inhomogeneous $\ell$-TASEP. Illustrations of EGGTART's accuracy through Monte Carlo simulations are provided in the Appendix.

\subsection*{The homogeneous $\ell$-TASEP}

\begin{figure}[h]
    \centering
    \includegraphics[width=.8\textwidth]{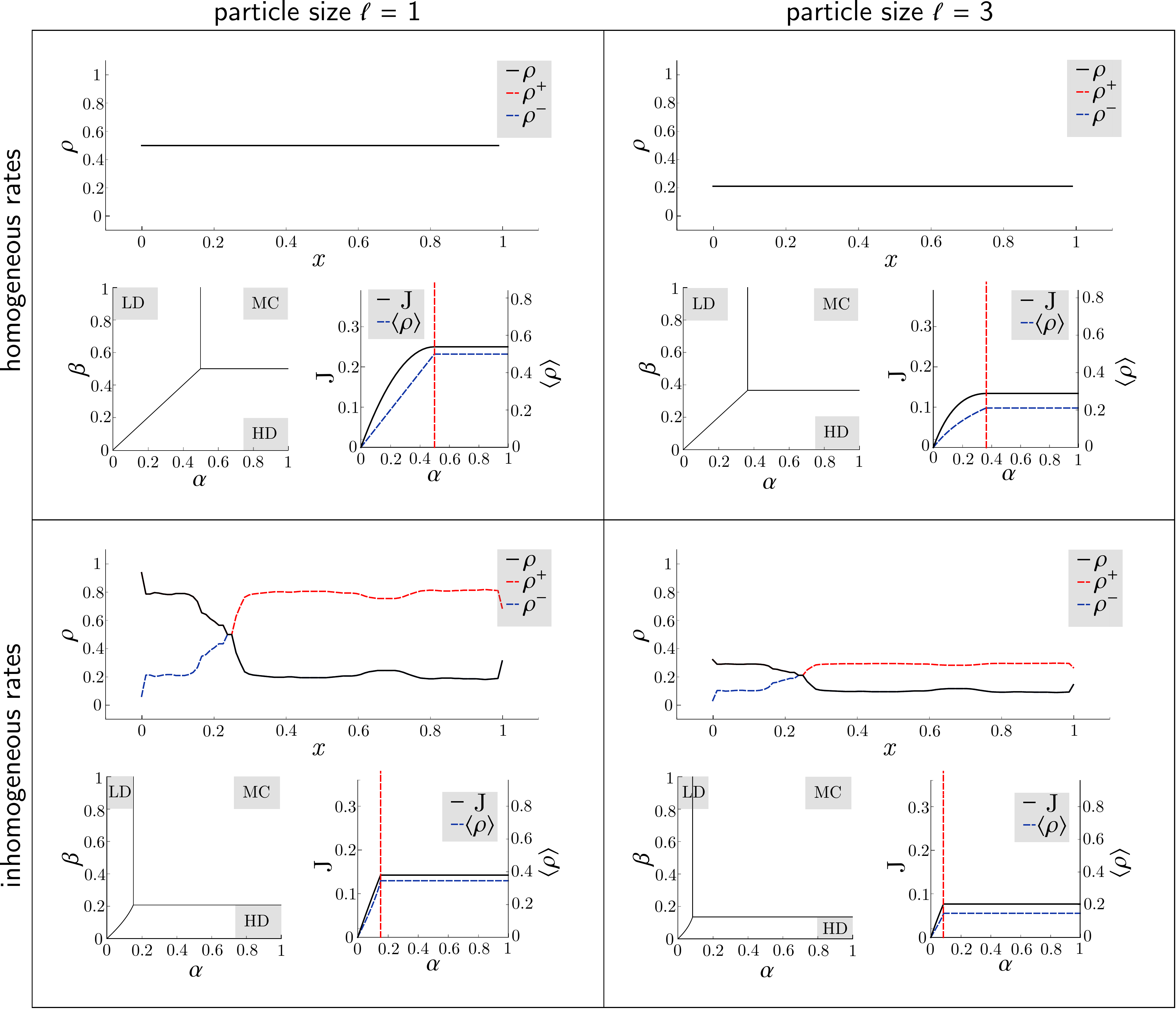}
    \caption{\textbf{The effect of inhomogeneous jump rates and extended particles.} The
    various impacts of increasing particle sizes and introducing
    inhomogeneity on the phase diagram, density profile and
    current can be displayed simultaneously using EGGTART. Particle sizes and rate
    functions can be adjusted independently allowing for investigation of
    individual (top right and bottom left) and joint effects (bottom right).
}
    \label{fig_table_figure} 
\end{figure}

The first TASEP model was introduced by MacDonald \emph{et al.} in \cite{macdonald1968} to model protein synthesis. Defined for homogeneous rates ($p_i \equiv 1$), its special case of $\ell = 1$ has been analytically solved through the Matrix Ansatz method \cite{derrida1993}. In the top left panel of \fref{fig_table_figure}, we show that EGGTART recovers the classical results obtained therein as special cases of the general model: 
Due to the symmetry of particles and holes, the resulting phase diagram is symmetric in $\alpha$ and $\beta$, with three regions (LD, HD and MC, see \textbf{Methods}) separated by simple lines intersecting at the critical point $(\alpha^*, \beta^*) = \left(1/2, 1/2 \right)$. 
Particle currents $J$ in LD, HD and MC are $\alpha(1-\alpha)$, $\beta(1-\beta)$ and $\frac{1}{4}$, respectively, with bulk densities $\rho$ given by $\alpha$, $\beta$ and $\frac{1}{2}$. 

General transport phenomena typically require particle size $\ell > 1$; 
e.g., 
ribosomes translating mRNA sequences occupy roughly 10 codons \cite{daoduc2018theoretical}. The $\ell$-TASEP for arbitrary $\ell \geq 1$ models this effect. 
As in the $1$-TASEP case, 
mean-field approaches can provide approximations to the phase diagram, currents and densities \cite{lakatos2003}, which are made precise 
in the hydrodynamic limit \cite{schonherr2005}. EGGTART also recovers these results, as seen in the top right panel of \fref{fig_table_figure}. 
The critical values $\alpha^*$ and $\beta^*$ shrink to $\frac{1}{1+\sqrt{\ell}}$, while $J$ decreases to $\frac{\alpha(1-\alpha)}{1+(\ell -1)\alpha}$, $\frac{\beta(1-\beta)}{1+(\ell -1)\beta}$ and $\frac{1}{(1+\sqrt{\ell})^2}$, in LD, HD and MC respectively (with similar adjustment for $\rho$), showing that increasing the size of particles leads to a global decrease of densities, currents and transport capacity. 

\subsection*{Intermediate cases: bottlenecks and linear rate function}

Simple lattice inhomogeneities have been studied extensively, as they promise closer approximation to actual transport phenomena while maintaining theoretical tractability. The most fundamental examples include single defects or defect clusters, i.e., sites or groups of sites with lower rates, that model ``slow'' codons along a mRNA sequence or structural imperfections of the structure or proteins in microtubular transport \cite{pierobon2006bottleneck, dong2007inhomogeneous}.
 Processes with monotonically-varying jump rates, which in the most elementary approach are modeled by linear functions, constitute another class of simple examples of inhomogeneity, and have notably
been observed in translation dynamics 
as the so-called 
  ``5' translational ramp'' \cite{daoduc2018impact}. EGGTART readily visualizes and quantifies such configurations, reproducing as before results previously reported in the literature. We provide more details on these intermediate examples and their corresponding biophysical applications in the SI file. 

\subsection*{The inhomogeneous $\ell$-TASEP}

The most general case treated by our model combines extended particles and heterogeneous rates. Such a generalization is for example necessary to properly model mRNA translation, as the local elongation rate of ribosomes (which occupy 10 codons) depends on multiple factors 
\cite{daoduc2018impact}. Therefore, most realistic modeling  of translation dynamics has been based on the inhomogeneous $\ell$-TASEP \cite{zur2016predictive}, with estimates of rates obtained using measurements of tRNA usage \cite{dana2012determinants} or, more recently, by analyzing ribosome profiling data \cite{daoduc2018impact}. Detailed analyses under such heterogeneous lattices have until recently been restricted to specific lattice configurations \cite{szavits2018deciphering}, with mean-field approximations suffering from numerical instabilities and imprecisions \cite{shaw2004mean}. 
With EGGTART we are able to directly and accurately quantify, and discern from each other, effects due to particle size and inhomogeneity: 
We observe
\textit{i)} the reduction in transport capacity and critical rates ($\alpha^*$ and $\beta^*$) associated with a decrease of the limiting jump rate $\lambda_{\min}$ (bottom left panel in \fref{fig_table_figure}) and extended particles (bottom right panel); \textit{ii)} deformation of the LD-HD phase separation (bottom panels)
; \textit{iii)} the branch switching of $\rho$ 
in MC (bottom panels); and \textit{iv)} the impact of $\lambda_0$ on the sensitivity of particle current to the initiation rate (see the SI file). Each of these principles has important consequences for optimizing ribosome usage and translation efficiency \cite{erdmann2020key}. Therefore, using the software to visualize and adjust these parameters can help practitioners understand a gene's evolutionary constraints or devise sequences most suitable for specific needs.

\section*{Discussion \& Conclusion}

Although the TASEP has been widely used and studied for several decades \cite{schadschneider2010, blythe2007nonequilibrium, zur2016predictive}, we provide here, to the best of our knowledge, the first computational tool for visualizing phase transitions, densities and fluxes of the TASEP in full generality, with an arbitrary particle size and inhomogeneous jump rates.  Our tool has several advantages, especially in the context of mRNA translation, compared to previous approaches mostly based on numerical simulations of biophysical models. While simulation studies struggle with high dimensional parameter spaces when exploring the determinants of transport efficiency and sensitivity, EGGTART allows one to quickly visualize the effects of the key system parameters using a graphical interface. 
As our theoretical result shows that particle flux and density depend on only a small set of parameters, 
EGGTART enables users to directly tune them and visualize their impact on the system. In particular, it allows to address key questions regarding transport efficiency \cite{erdmann2020key}: How far is the system from its transport capacity? What are the consequences of modifying jump rates (to account for defects, mutations, decrease of the particle pool etc.)? Is the 
system well optimized and to which parameters is it most sensitive? In the context of gene expression, these questions are typically asked on a gene by gene level, so a computational tool for immediate quantification and fine tuning are essential. We thus believe that EGGTART will aid as a powerful tool for studying biophysical transport systems, both theoretically and in relation to interpreting experimental data (e.g., local rates inferred from ribosome profiling data \cite{daoduc2018impact}). 

\section*{Acknowledgments}
This research is supported in part by an NIH grant R35-GM134922. 
Y.S.S. is a Chan Zuckerberg Biohub Investigator.

\newpage
\section*{Appendix}
\appendix

\section{Illustration of EGGTART on Examples of Intermediate Complexity}

Two popular and well-studied special cases of the general inhomogeneous $\ell$-TASEP are lattices with defect clusters and linear jumping rates. Here we detail how EGGTART recovers previously known results in these two examples, while allowing exploration of closely related models that so far had proven inaccessible.

\subsection*{Bottleneck induced by defect sites}

Due to its relevance in applications and mathematical tractability, many studies of the TASEP have considered inhomogeneity created by defects, i.e.,  sites with slower rates \cite{korkmazhan2017dynamics, dong2007inhomogeneous,  pierobon2006bottleneck}. For example, defects can be associated  with ``slow'' codons that  potentially limit the 
protein synthesis rate during mRNA translation  \cite{korkmazhan2017dynamics, dong2007inhomogeneous}, or with structural imperfections of the microtubular structure or 
motor proteins transported along microtubules  \cite{pierobon2006bottleneck, goldstein2001kinesin}.
We illustrate the simplest case of a single defect in the top three panels of  \fref{fig_defect}, before demonstrating the behavior of two disjoint defects in the bottom two panels. In 
the former case, mean-field approximations successfully approximate the phase diagram and density profiles \cite{shaw2004mean, kolomeisky1998phase, 
dong2007inhomogeneous, basu2017invariant, greulich2008phase}. Compared with the homogeneous system, the main effects caused by a single defect are an enlarged MC phase region with a decrease of the  maximal current, as well as significantly altered density profiles. LD and HD densities exhibit local deviations around the defect site, while in MC the defect site  acts as a separator between a region of high density on the left and a region of low  density on the right. For macroscopic clusters of defect sites, refined mean-field approaches have shown that  these effects persist \cite{chou2004clustered, liu2010synchronous}. Using the hydrodynamic  limit, we are able to recover these results precisely:  Defining $\lambda(x) = 1 - \eta_{\varepsilon}(x)$, where $\eta_{\varepsilon}$ is a  suitably normalized bump function centered around the bottleneck $x_0$, we find that with  $
\lambda_{\min} = \lambda(x_0) < 1$,
\begin{align}
    \alpha^{\ast} = \beta^{\ast} & = \dfrac{1}{1+\sqrt{\ell}}\bigg( 1- \sqrt{ 1-\lambda_{\min} } \bigg), \nonumber \\
      J_{\max} & = \dfrac{\lambda_{\min}}{(1+\sqrt{\ell})^2},
    \label{eq:defectalpha}
\end{align} 
leading to the reduction in transport capacity, shifts in the phase diagram  and local density perturbations (top two panels in \fref{fig_defect}) outlined in the main manuscript. The co-existence 
of low and high density regions in the MC phase is reflected in the branch switching phenomenon (middle panel of \fref{fig_defect}). In addition, EGGTART allows to interactively
explore the dynamics associated with the emergence of a second defect located downstream the first one (with respective rates $\lambda^{(2)}$ and $\lambda^{(1)}$). In the MC regime, an initial local distortion of $\rho$ around the second bottleneck for $\lambda^{(1)} < \lambda^{(2)}$ turns into a global distortion as soon as $\lambda^{(2)}<\lambda^{(1)}$, as shown in the bottom two panels of \fref{fig_defect}.


\subsection*{Linear rate function}

Monotonically-varying jump rates are another simple example of spatial inhomogeneity.  
This kind of pattern has been observed in translation dynamics, where
the mean ribosome elongation rate, obtained by averaging over all mRNA transcripts, 
increases between codon positions $\sim 50$ and $200$, 
leading to the so-called 
``5' translational ramp'' \cite{daoduc2018impact}. 
Such variation can be modeled to first approximation by a linear rate function 
$\lambda(x) = s(x-1) + 1$ for $s\in [0,1]$ (the case of decreasing rates with $s<0$ 
can be treated analogously). With $\lambda_{\mathrm{min}} = \lambda_0 = 1-s$ and 
$\lambda_1 = 1$, we can easily compute $\alpha^{\ast}$ and $\beta^{\ast}$ using the formulae obtained in \cite{erdmann2020key}, obtaining for the particular case of $\ell=1$,
\begin{equation}
    \alpha^{\ast} = \dfrac{1-s}{2} \hspace{4mm}\text{and}\hspace{4mm} \beta^{\ast} = \dfrac{1}{2}\bigg( 1 -
    \sqrt{s} \bigg),
    \label{eq:linearstars}
\end{equation}
reflecting the asymmetry of the model. Similarly, the phase boundary between LD and HD is explicitly given by the curve
\begin{equation}
    \alpha(1-\alpha-s) = \beta(1-\beta)(1-s),
    \label{eq:HDLDtransition}
\end{equation}
recovering previous results obtained through mean-field approximations 
\cite{stinchcombe2011smoothly}. Notably, this curve is non-linear (see right half of \suppfref{fig_linear}), again evidencing the asymmetry of the model.
Since $\lambda$ achieves its minimum at the lattice 
entry site, both LD and MC profiles are described by the lower branch (\suppfref{fig_linear} top and bottom panel), allowing for densities larger than $\rho^*= (\ell+\sqrt{\ell})^{-1}$ only in HD (\suppfref{fig_linear} middle panel). Lastly, we highlight a robustness property of this linear model, which so far has eluded attention: Lattices with linearly increasing jump rates belong to a family of systems (namely, those for which $\lambda_0 = \lambda_{\min}$) whose currents are \textit{maximally insensitive} to fluctuations in the initiation rate $\alpha$. This follows directly from our explicit description of the particle current $J_L$ as a function of $\lambda_0, \lambda_{\min}$ and $\alpha$, and is illustrated in \fref{fig_sensitivity}. This description only became available through consideration of the full general inhomogeneous $\ell$-TASEP, explaining why the role of $\lambda_0$ had remained unnoticed.

Robustness characterizations of particle currents like these, together with sensitivity constraints on particle densities, as illustrated in the bottom two panels of \fref{fig_defect}, have important consequences for optimizing ribosome usage and translation efficiency \cite{erdmann2020key}. By allowing for immediate quantification of such phenomena, EGGTART enables users to assess the optimality of any given system and, if necessary, open avenues to improve its efficiency.

\section{Validation of EGGTART's accuracy}
A detailed demonstration of the convergence speed associated with the hydrodynamic limit (and therefore, its accuracy when describing finite systems) has been carried out in \cite{erdmann2020key}. As shown there, convergence is generally fast, with lattices as short as $100$ sites typically producing particle currents and densities indiscernible from our hydrodynamic predictions. \fref{fig_supp_simu} illustrates this accuracy by comparing empirical samples of Monte-Carlo simulations to the theoretical profiles provided by EGGTART.

\bibliographystyle{unsrt}
\bibliography{eggtart}
\newpage
\section{Supplementary Figures}

\renewcommand{\thefigure}{A\arabic{figure}}
\setcounter{figure}{0} 

\begin{figure}[h!]
     \centering
     \includegraphics[width=.8\textwidth]{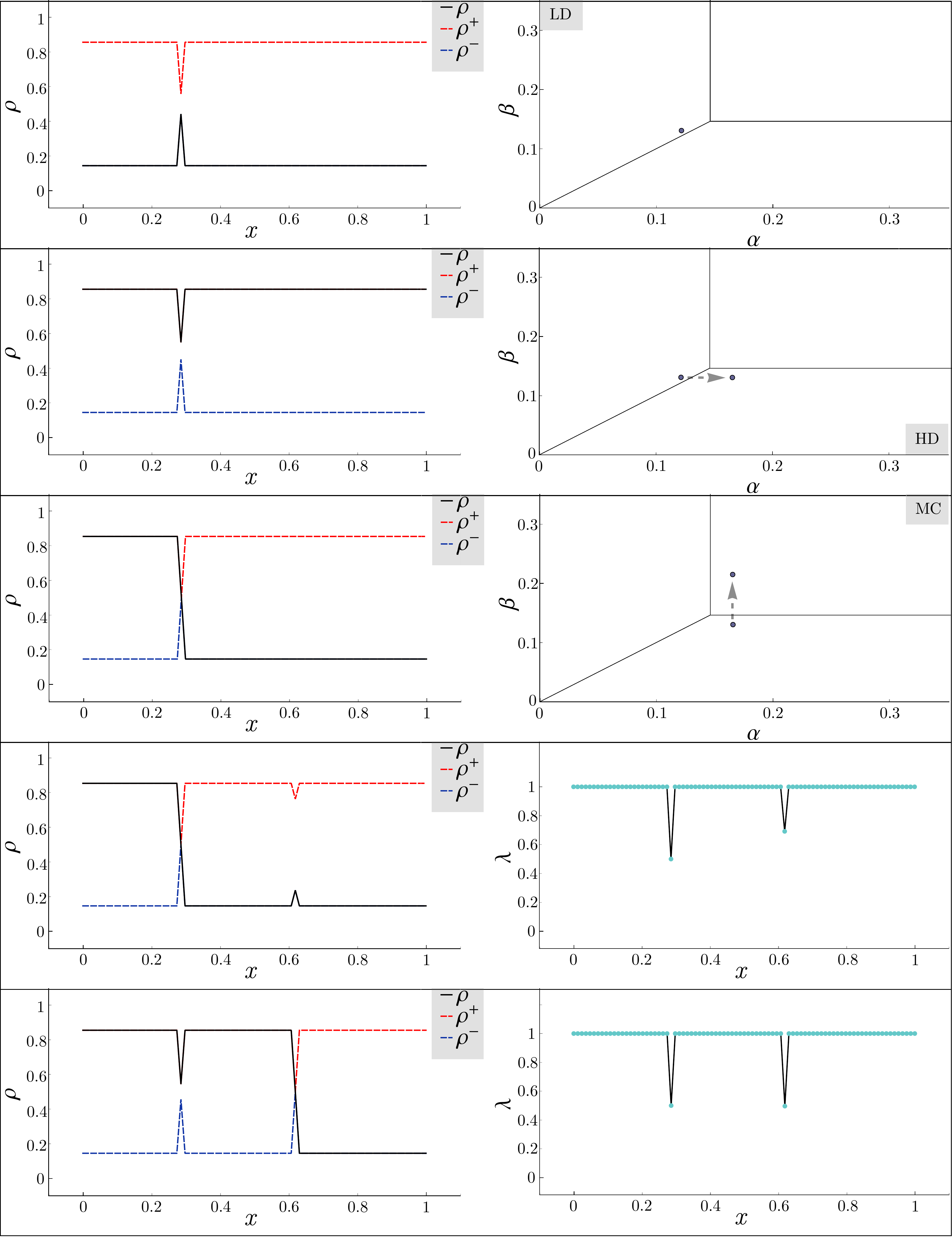}
     \caption{\textbf{Interactive visualization on the example of defect clusters.} EGGTART
     allows for convenient exploration of discontinuous phenomena around phase
     transitions (top three panels) and singular rate configurations (bottom two
     panels).}
     \label{fig_defect} 
 \end{figure}
 \newpage
 
\begin{figure}[h!]
    \centering
    \includegraphics[width=\textwidth]{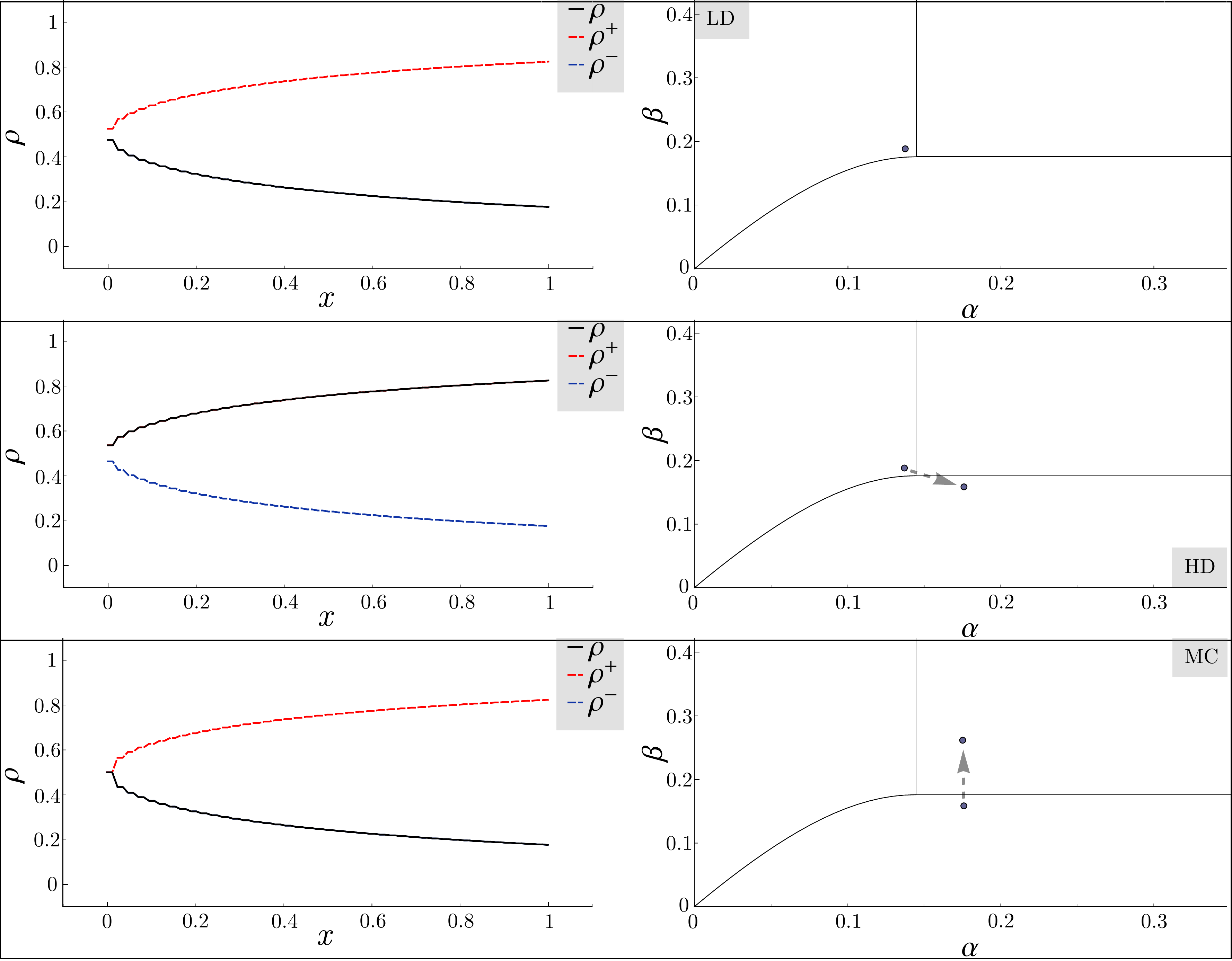}
    \caption{\textbf{Interactive visualization on the example of linear rate functions.} EGGTART
        correctly reproduces singular cases of the inhomogeneous $\ell$-TASEP, when either
        $\lambda_0 = \lambda_{\min}$ (plotted here) or $\lambda_1 = \lambda_{\min}$, and
        LD and MC densities coincide. 
    }
    \label{fig_linear}
\end{figure}
\newpage

\begin{figure}[h!]
    \centering
    \includegraphics[width=\textwidth]{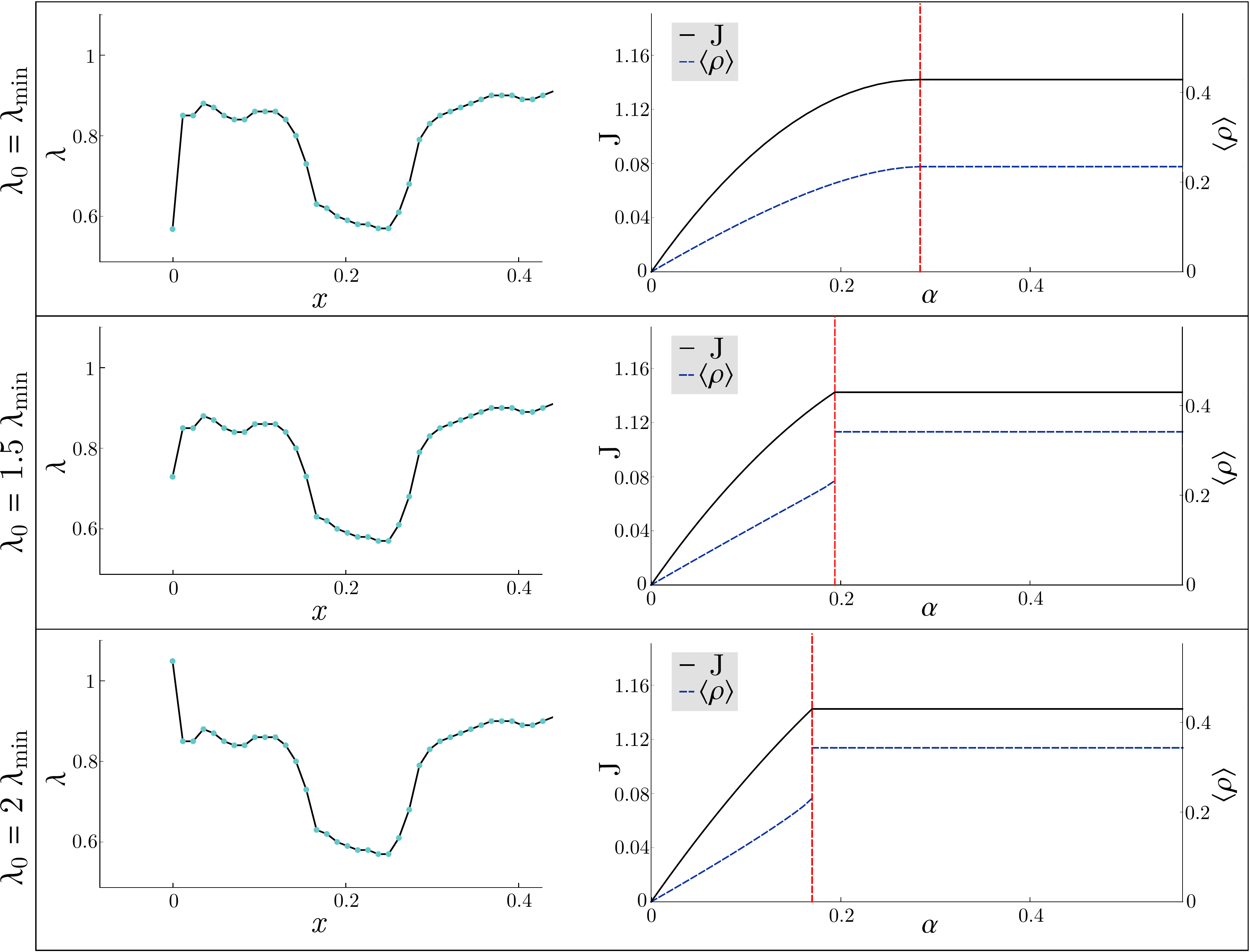}
    \caption{\textbf{Impact of $\lambda_0$ on sensitivity.} The precise effect of changes
        in the initial rate $\lambda_0$ on current $J$ and mean density $\langle \rho \rangle_x = \int_{0}^{1} \rho(x) \ \mathrm{d}x$ can be easily visualized using the interactive interface of
        EGGTART: Larger $\lambda_0$ lead to higher sensitivity of $J$ to changes
        in $\alpha$, and quicker saturation at maximum capacity.
        Moreover, $\langle \rho \rangle_x$ phase transitions are present only if
        $\lambda_0 > \lambda_{\min}$.
}
    \label{fig_sensitivity}
\end{figure}
\newpage

\begin{figure}[h!]
    \centering
    \includegraphics[width=\textwidth]{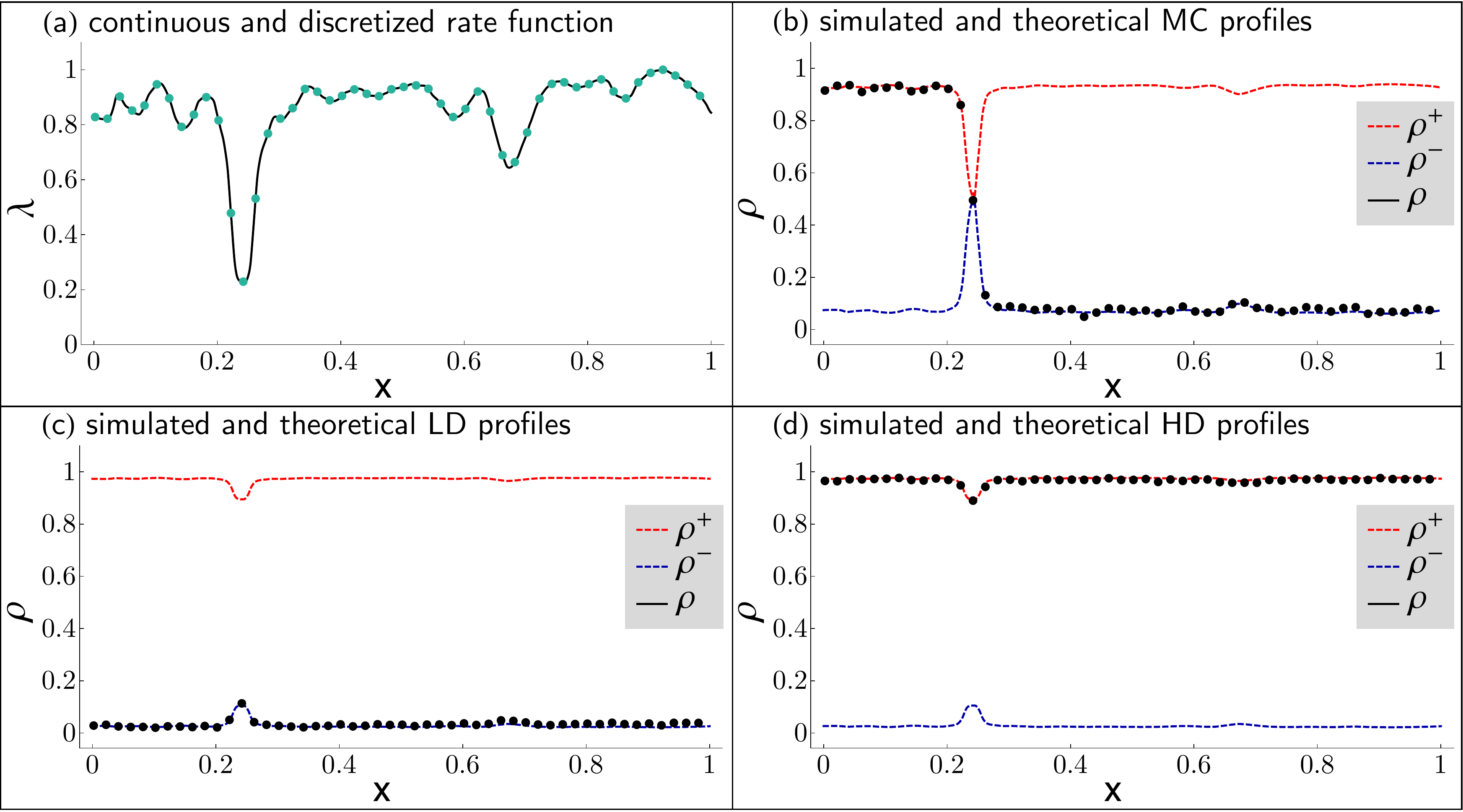}
    \caption{\textbf{Comparison with Monte Carlo simulations.} $5\times 10^7$ Monte-Carlo
    iterations were averaged after $10^7$ burn-in steps on a lattice of size
    $100$, $\ell = 1$ particles and the rate function given in panel (a). The
    resulting simulated density profiles (in dots) agree well with our theoretical
    predictions in all regimes of the phase diagram, obtained by adjusting $\alpha$ and $\beta$.}
    \label{fig_supp_simu} 
\end{figure}

\end{document}